\documentclass[aps,amssymb,prl,twocolumn,showpacs]{revtex4}
\usepackage{epsfig}
\usepackage{array}

\def\al{{\it et. al.} }

\def\0{\varnothing}

\begin{document}
\title{Novel phase-separation transition in one-dimensional driven models}
\author{Y. Kafri$^{1}$, E. Levine$^{1}$, D. Mukamel$^{1}$, G. M.
Sch\"{u}tz$^{2}$, and R. D. Willmann$^{1,2}$}
\affiliation{$^1$Department of Physics of Complex Systems,
Weizmann Institute of Science, Rehovot, Israel 76100.\\
$^2$Institut f\"{u}r Festk\"{o}rperforschung, Forschungszentrum
J\"{u}lich, 52425 J\"{u}lich, Germany.}
\date{\today}

\begin{abstract}
A class of models of driven diffusive systems which is shown to
exhibit phase separation in $d=1$ dimensions is introduced. Unlike
all previously studied models exhibiting similar phenomena, here
the phase separated state is fluctuating in the bulk of the
macroscopic domains. The nature of the phase transition from the
homogeneous to the phase separated state is discussed in view of a recently introduced
criterion for phase separation in one-dimensional driven systems.
\end{abstract}

\maketitle

One dimensional driven diffusive systems have attracted
considerable attention in recent years \cite{Mukamel00}. It has
been demonstrated in numerous studies that unlike systems in
thermal equilibrium, certain driven diffusive models with local
noisy dynamics do exhibit phenomena like phase transitions and
phase separation. More recently a general criterion for the
existence of phase separation in driven one dimensional models has
been introduced \cite{Conj}. The criterion relates the existence
of phase separation in a given model to the rate at which domains
of various sizes exchange particles with each other. Assuming that
for a domain of length $n$ this rate is given by the steady state
current $J_n$ which flows through it, phase separation was
suggested to exist only in the following cases:
%\begin{description}
%\item[Case A.]
either the current vanishes in the thermodynamic limit,
\begin{equation}
 J_n \to 0\;\; \mbox{as}\;\; n \to \infty \hspace{1.2cm}(\mbox{\bf
Case A})\; \label{eq:caseA}
\end{equation}
%\item[Case B.]
or the behavior of the current for large domains is of the form
\begin{equation}
 J_n \sim J_\infty\left(1+b/n^{\sigma}\right)
\hspace{1.0cm} (\mbox{\bf Case B}) \label{eq:caseB}
\end{equation}
for either $\sigma<1$ and $b>0$, or for $\sigma=1$ and $b>2$.

The nature of the phase separated states is rather different in
the two cases. In Case A the phase separated states were found to
be of a rather simple nature, characterized by coexistence of pure
domains, each consisting of a single type of particles. Thus, the
particle density in the interior of a domain is non-fluctuating.
Density fluctuations are limited to finite regions around the
domain boundaries. Such steady states were termed {\em strongly}
phase separated. Moreover, in this case phase separation is
expected to take place at {\em any} density, no matter how small.
On the other hand, in Case B the phase separated state is expected
to be fluctuating in the bulk of the macroscopic domains, as is
normally expected in a noisy system. It exists only at high enough
densities, while at low densities the system is homogeneous. This
phase was termed {\it condensed} as the mechanism of the
transition is similar to that of the Bose-Einstein condensation.

So far, all one-dimensional models found to phase separate are of
type A \cite{Evans98,Rittenberg99,LBR00}, and thus they exhibit
strong phase separation at any density. In these models more than
one species of particles is involved. In a recent study by Arndt
\al (AHR)~\cite{Rittenberg99} an interesting two species driven
model was introduced. It was suggested, based on numerical
simulations, that the model exhibits a condensed phase separated
state, whereby the particle densities fluctuate in the interior of
the coexisting domains, and not just at the domain boundaries. In
this state, a region with a high density of particles of both
species coexists with a low density region. Moreover, the model
has non-vanishing currents even in the thermodynamic limit. As in
equilibrium phase separation it has been suggested that this state
exists only at sufficiently high densities. However, a subsequent
exact solution of the model \cite{Sasamoto00} shows that what
numerically seems like a condensed state is in fact homogeneous,
with a very large but finite correlation length. Further analysis
of this model, in the light of the criterion suggested in
\cite{Conj} shows that the currents $J_n$ corresponding to this
model are given by the form B, with $\sigma=1$ and $b=3/2$
\cite{Conj}. Therefore, according to the criterion, no phase
separation takes place.

Another example of a model which was suggested to exhibit phase
separation into a fluctuating macroscopically inhomogeneous state
is the two-lane model introduced by Korniss \al \cite{Zia}. While
numerical studies of the model indicate that such a phase exists
in the model, studies of the current $J_n$ of finite domains
suggests that it is of type B with $\sigma=1$ and $b \simeq 0.8$
\cite{Conj}, indicating, again, that no phase separation exists in
this model. Thus the question of whether a phase separation of
type B exists remains an intriguing open question.

In this Letter we introduce a class of models which are
demonstrated to be of type B, with $\sigma=1$ and $b>2$. According
to the criterion conjectured in \cite{Conj} this class is expected
to exhibit a phase transition to a phase separated {\em condensed}
state. Thus at high densities these models exhibit a novel phase
separation with non-vanishing currents in the thermodynamic limit,
and bulk fluctuations which are not restricted to the vicinity of
the domain boundaries. To our knowledge, this is the first example
of a genuine transition of this type in one-dimensional driven
systems.

We now define this class of models in detail. We consider a
one-dimensional ring with $L$ sites. Each site $i$ can be either
vacant ($0$) or occupied by a positive ($+$) or a negative ($-$)
particle (or charge). Positive particles are driven to the right
while negative particles are driven to the left. In addition to
the hard-core repulsion, particles are subject to short-range
interactions. These interactions are ``ferromagnetic'', in the
sense that particles of the same kind attract each other. The
dynamics conserves the number of particles of each species, $N_+$
and $N_-$. The total density of particles in the system is
$\rho=(N_++N_-)/L$. The model is defined by a random-sequential
local dynamics, whereby a pair of nearest-neighbor sites is
selected at random, and the particles are exchanged with the
following rates:
\begin{equation}
\begin{array}{cccl}
+- &\to& -+ &\hspace{0.5cm}\mbox{with rate }\hspace{0.5cm} 1 - \Delta H \\
+\,0 &\to& 0\,+ &\hspace{0.5cm} \mbox{with rate }\hspace{0.5cm} \alpha \\
0\,- &\to& -\,0 &\hspace{0.5cm} \mbox{with rate }\hspace{0.5cm}
\alpha \;.
\end{array}
\label{eq:rates}
\end{equation}
Here $\Delta H$ is the difference in the ferromagnetic
interactions between the final and the initial configurations. We
begin by considering a model with only nearest neighbor
interactions,
\begin{equation}
H=-\epsilon/4 \sum_i{s_is_{i+1}}\;. \label{eq:H}
\end{equation}
Here $s_i=+1$ ($-1$) if site $i$ is occupied by a $+$ ($-$)
particle, and $s_i=0$ if site $i$ is vacant. The interaction
parameter $\epsilon$ satisfies $0 \leq \epsilon < 1$ to ensure
positive transition rates. The model is a generalization of the
Katz-Lebowitz-Spohn (KLS) model, introduced in \cite{KLS} and
studied in detail in \cite{Hager01}, in which the lattice is fully
occupied by charges and no vacancies exist. In this Letter we
consider the case where the number of positive and negative
particles is equal, $N_+=N_-$.

\begin{figure}
\centerline{\epsfig{file=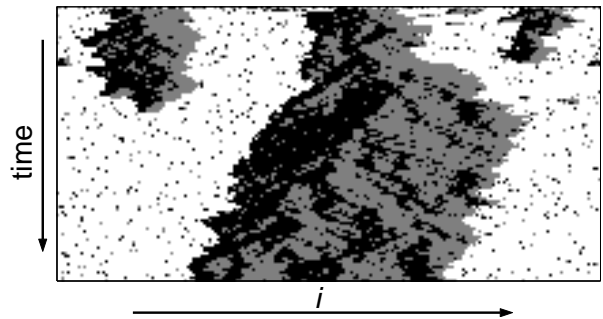,width=8truecm}}
\caption{\label{fig:config} Evolution of a random initial
configuration of model (\ref{eq:rates}) with nearest-neighbor
interactions, on a ring of $200$ sites. Here $\epsilon=0.9$,
$\alpha=2$, and the particle density is $\rho=0.5$. Positive
particles are colored black, and negative particles are colored
grey. One hundred snapshots of the system are shown every 100
Monte-Carlo sweeps.}
\end{figure}

We will demonstrate that for a certain range of the parameters
defining the dynamics, namely for $\epsilon > 0.8$ and
sufficiently large $\alpha$ (to be discussed below), a phase
separation transition occurs as the density $\rho$ is increased
above a critical density $\rho_c$. In the phase separated state a
macroscopic domain, composed of positive and negative particles,
coexists with a fluid phase, which consists of small domains of
particles (of both charges) separated by vacancies. Typical
configurations obtained during the time evolution of the model
starting from a random initial configuration are given in Fig.
\ref{fig:config}. This figure suggests that a coarsening process
takes place, leading to a phase separated state as described
above. However, this by itself cannot be interpreted as a
demonstration of phase separation in these models. The reason is
that this behavior may very well be a result of a very large but
finite correlation length, as is the case in the AHR
\cite{Rittenberg99,Sasamoto00} and the two-lane \cite{Zia} models
discussed above \cite{Sharp}. We thus apply the criterion \cite{Conj} in order to
analyze the possible existence of phase separation in this model.

To this end we note that a domain may be defined as an
uninterrupted sequence of positive and negative particles bounded
by vacancies from both ends. The current $J_n$ corresponding to
such a domain of length $n$ may thus be determined by studying an
open chain, fully occupied by positive and negative particles,
with entrance and exit rates $\alpha$. This is just the
one-dimensional KLS model on an open chain. Phase separation is
expected to take place only for sufficiently large $\alpha$. We
consider $\alpha$ such that the system is in its maximal current
state, whereby $J_\infty$ assumes its maximum possible value, and
is independent of $\alpha$.

To evaluate $J_n$ we first consider the KLS model on a ring of $n$
sites with no vacancies. We then extend these results to study the
behavior of an open chain. Since we are interested in the maximal
current phase we consider equal number of positive and negative
particles $n_+=n_-=n/2$. It can be shown, as was done for the
noisy Burger's equation \cite{Krug90,Krug97}, that under quite
general conditions, to be discussed below, the current $J_n$ takes
the following form for large $n$,
\begin{equation}
\label{eq:krug} J_n=J_\infty \left( 1 - \; \frac{\lambda \;
\kappa}{2 J_\infty} \frac{1}{n} \right) \;.
\end{equation}
Here $\lambda= \partial^2 J_\infty / \partial \rho_+^2$ is the
second derivative of the current with respect to the density of
positive particles $\rho_+$ in the system.  The compressibility
analog $\kappa$ is defined by $\kappa=\lim_{n \to \infty} n^{-1}
\left (\langle n_+^2 \rangle - \langle n_+ \rangle^2 \right)$, as
calculated within a grand canonical ensemble, as explained below.
This can be demonstrated by considering the current $J_n(n_+)$ for
charge densities close to $n_+=n_-=n/2$. Expanding $J_n(n_+)$ in
powers of $\Delta n_+ = n_+ - n/2$ one has
\begin{equation}
J_n(n_+) = J_n (n/2)+ J_n'\;\Delta n_+ + \frac{1}{2}J_n''\;(\Delta
n_+)^2 \label{eq:expand}
\end{equation}
where the derivatives $J_n'$ and $J_n''$ are taken with respect to
$n_+$ and evaluated at $n/2$. We average (\ref{eq:expand}) over
$n_+$ with the steady state weights of a grand canonical ensemble.
This is done by introducing a chemical potential $\mu$ which
ensures that the average density satisfies $\langle n_+ \rangle=
n/2$. We find
\begin{equation}
\langle J_n(n_+) \rangle_\mu = J_n (n/2)+ \frac{1}{2}
J_n''\;\langle (\Delta n_+)^2 \rangle_\mu \;.\label{eq:expandavg}
\end{equation}
Noting that $\langle J_n(n_+) \rangle_\mu$ is $J_\infty$ in the $n
\to \infty$ limit, and $J_n(n/2)$ is just $J_n$, Eq. \ref{eq:krug}
is obtained. Here we made use of the fact that finite size
corrections to $\langle J_n(n_+) \rangle_\mu$, resulting from the
next to leading eigenvalue of the transfer-matrix of the
steady-state distribution, are exponentially small in $n$ and may
thus be neglected. The result of Eq.~\ref{eq:krug} is rather
general, and is independent of the exact form of the steady-state
particle distribution. This is provided that the weights of the
microscopic configurations are local and thus the density and
chemical potential ensembles are equivalent.

In fact, an alternative way to derive (\ref{eq:krug}) is to
consider the correspondence between the driven lattice-gas models
and the noisy Burger's equation or the Kardar-Parisi-Zhang (KPZ)
equation for interface growth in $1+1$ dimensions \cite{BKS85}. In
these models $J_n$ corresponds to the growth velocity of the
interface. Eq. \ref{eq:krug} has been derived in
\cite{Krug90,Krug97}, where $\lambda$ is the coefficient of the
non-linear term in the KPZ equation. The equivalence of the two
alternative approaches relies on the fact that both $\kappa$ and
$\lambda$ are invariant under renormalization transformations.

The result (\ref{eq:krug}) can be used to evaluate $J_n$ for the
KLS model. It has been shown \cite{KLS,Hager01} that for a ring
geometry the steady state weight of a configuration
$\left\{\tau_i\right\}$ is
\begin{equation}
P(\left\{\tau_i\right\}) = e^{- \beta {\cal H}}\;\;;\;\;{\cal
H}=-\sum_{i=1}^{n}{\tau_i \tau_{i+1}} - \mu
\sum_{i=1}^{n}{\tau_i}\;, \label{eq:ising}
\end{equation}
with $\tau_i=\pm 1$ for positive and negative charges
respectively, $e^{4\beta}=(1-\epsilon)/(1+\epsilon)$, and $\mu$
serves as a chemical potential which controls the density of, say,
the positive particles. The chemical potential $\mu$ vanishes for
the case $n_+=n_-$. Using (\ref{eq:ising}) expressions for
$\kappa(\epsilon)$ and $J_\infty(\epsilon)$ of this model have
been obtained in \cite{Hager01}.

\begin{figure}
\centerline{\epsfig{file=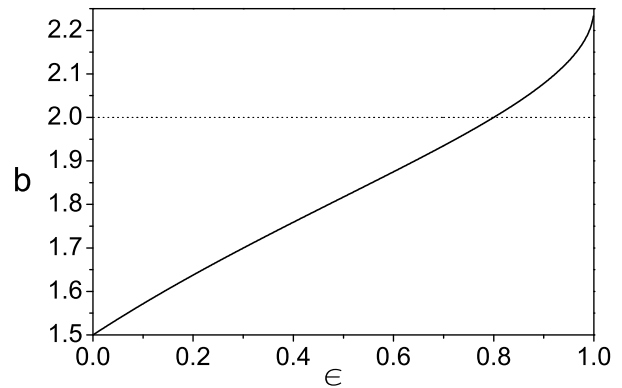,width=8truecm}}
\caption{\label{fig:bofe} The coefficient $b(\epsilon$), Eq.
\ref{eq:bofe}.}
\end{figure}

We now consider the KLS model in an open chain, which is the
relevant geometry in applying the phase-separation criterion. It
has been argued \cite{Krug90} that the finite size correction to
the current of an open chain is given by the corresponding
correction in a ring geometry, up to a universal multiplicative
constant $c$ which depends only on the boundary conditions. In the
maximal current phase, $c$ was found to be $3/2$. Thus the current
of an open system is given by (\ref{eq:caseB}) with $\sigma=1$ and
\begin{equation}
b(\epsilon)=-c \frac{\lambda(\epsilon) \kappa(\epsilon)}{2
J_\infty(\epsilon)}\;.
\end{equation}
Using the values of $J_\infty$ and $\kappa$ obtained in
\cite{Hager01} and $c=3/2$ we find
\begin{equation}
b(\epsilon)=\frac{3}{2}\;\frac{(2+\epsilon)\upsilon+2\epsilon}{2(\upsilon+\epsilon)}\;;\;
\upsilon=\sqrt{\frac{1+\epsilon}{1-\epsilon}}+1\;. \label{eq:bofe}
\end{equation}
In figure \ref{fig:bofe} the coefficient $b(\epsilon)$ is plotted
for $0 \leq \epsilon < 1$. This curve has been verified by direct
numerical simulations of the KLS model on an open chain in the
maximal current phase, demonstrating that the prefactor $c$ indeed
does not depend on $\epsilon$ .
%It is known that $b(0)=3/2$ in the maximal current phase
%\cite{TASEP}, which yields $c=3/2$ in agreement with
%\cite{Krug90}.
Using (\ref{eq:bofe}) it is readily seen that for $\epsilon > 0.8$
the value of $b$ is larger than $2$.

According to the criterion conjectured in \cite{Conj} one expects
phase separation to take place at high densities in model
(\ref{eq:rates}) for $\epsilon > 0.8$, as long as $\alpha$ is such
that the KLS model is in the maximal current phase. This {\em
condensed} phase separated state belongs to Case B of the
criterion. We have carried out extensive numerical simulations of
the dynamics of the model for various values of $\epsilon$. We
find that for $\epsilon \lesssim 0.4$ no phase separation is
observed. However, for $\epsilon > 0.4$ simulation of systems of
sizes up to $L=10^6$ show that the system evolves towards what
seems to be a phase separated state at sufficiently large
densities. We argue that a genuine phase separation takes place
only for $\epsilon > 0.8$. On the other hand, the seemingly phase
separation found in simulations for $0.4 \lesssim \epsilon < 0.8$,
is only a result of large but finite correlation lengths, as was
found in the AHR and in the two-lane models. As pointed out in
\cite{Sharp} such a behavior is related to corrections of order
$1/n^2$ and higher in the current (\ref{eq:caseB}). These
corrections were shown to lead to a crossover with a very sharp
increase in the correlation length, which could be erroneously
interpreted as a genuine phase transition in numerical studies of
finite systems.

We now discuss the phase transition leading to the phase separated
state. According to \cite{Conj} the domain size distribution just
below the transition takes the form
\begin{equation}
{\cal P}(n) \sim \frac{1}{n^b}e^{-n/\xi}
\end{equation}
where $\xi$ is the correlation length, which diverges at the
transition. The particle density in the system is related to $\xi$
by $\rho/(1-\rho) = \sum{n{\cal P}(n)}/\sum{{\cal P}(n)}$. The
critical density $\rho_c$ is given by this expression with $\xi
\to \infty$. Note that with this form of the distribution
function, $\rho_c$ is $1$ in the limit $b \searrow 2$, and is a
decreasing function of $b$. It is straightforward to show
\cite{DNA} that the divergence of the correlation length at the
critical density is given by
\begin{equation}
\xi \sim \left\{ \begin{array}{lcl} |\rho-\rho_c|^{-\frac{1}{b-2}}
&,& 2<b<3 \nonumber\\ \nonumber \\|\rho-\rho_c|^{-1} &,& b>3\;.
\end{array}\right.
\end{equation}
It is worthwhile noting that while $\partial \xi^{-1}/\partial
\rho$ is continuous at the transition for $2<b<3$, it exhibits a
discontinuity for $b>3$. The transition may thus be considered
continuous for $2<b<3$ and first-order for $b>3$.

\begin{figure}
\centerline{\epsfig{file=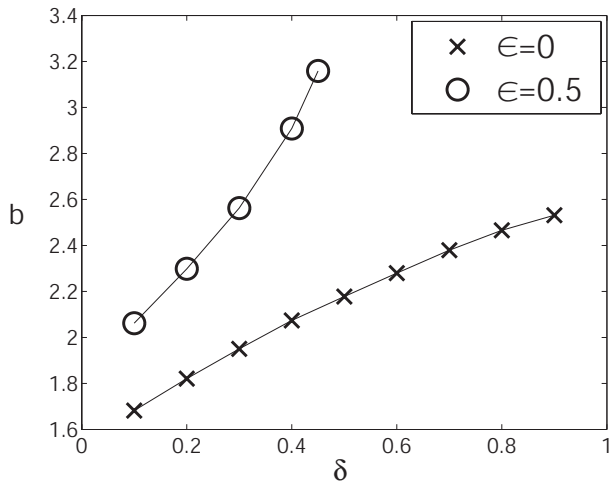,width=8truecm}}
\caption{\label{fig:bofdelta} The coefficient $b(\delta$), as
calculated from Monte-Carlo simulations of domains of sizes up to
$1024$. Data are shown for $\epsilon=0$ and $\epsilon=0.5$ .}
\end{figure}

In the model defined above $b$ is found to satisfy $3/2 \leq b <
9/4$. It is natural to ask whether larger values of $b$ could be
reached by increasing the range of the interactions. To answer
this question we have extended model (\ref{eq:H}) to include
next-nearest-neighbor interactions as well, and consider
\begin{equation}
H=-\epsilon/4 \sum_i{s_is_{i+1}}-\delta/4 \sum_i{s_is_{i+2}}\;.
\label{eq:hdelta}
\end{equation}
We have calculated the value of $b$ as a function of $\delta$ by
Monte-Carlo simulations. This is done by measuring the current
$J_n$ in an open system of size $n$, which is fully occupied by
positive and negative particles. At the boundaries, the coupling
to the rest of the system is modeled by injection of positive
(negative) particles with rates $\alpha$ at the left (right).
Simulating systems of size up to $1024$ enables us to fit the
measured values of $J_n$ to the form (\ref{eq:caseB}) with
$\sigma=1$, and to extract $b$. In figure \ref{fig:bofdelta} we
plot $b$ as a function of $\delta$, for $\epsilon=0$ and for
$\epsilon=0.5$. We find that by extending the range of the
interactions one can increase $b$ to values even larger than $3$,
where the phase separation transition is expected to be first
order.

In summary, a class of driven diffusive models in one-dimension is
introduced and analyzed using a recently conjectured criterion for
phase separation \cite{Conj}. These models are shown to exhibit a
novel type of phase separation. In the phase separated state of
these models the density is fluctuating in the bulk of the
domains. Moreover, the models exhibit a homogeneous state at low
densities, and a phase transition into the phase separated state
occurs at a critical density. The nature of the phase transition
in these models is also discussed.

While the validity of the criterion was proved for the AHR model,
its general validity was conjectured based on some plausible
assumptions on the behavior of the coarsening domains \cite{Conj}.
It would be of interest to analyze the class of models introduced
in the present study by other analytical means, in order to verify
the validity of the criterion.

\begin{acknowledgments}
We thank D. Kandel and G. Ziv for comments. The support of the
Israeli Science Foundation and the Einstein Center are gratefully
acknowledged. RDW thanks the Minerva Foundation for support
during his six-month stay at the Weizmann Institute.

\end{acknowledgments}

\end{document}